\documentclass[11pt,letterpaper,abstract=on]{scrartcl}

\usepackage[nochapters]{classicthesis}
\usepackage[left=1in,right=1in,top=1in,bottom=1in]{geometry}
\usepackage[affil-it]{authblk}

\usepackage[utf8]{inputenc}
\usepackage[T1]{fontenc}
\usepackage{libertine}
\usepackage{microtype}
\usepackage{graphicx}
\newcommand{\ie}{{i.\,e.}}

\newcommand{\eg}{{e.\,g.}}

\newcommand{\spara}[1]{\smallskip\noindent{\textbf{#1}.}}

\date{}

\title{\Large The Effect of Pok\'emon Go on The Pulse of the City: \\ A Natural Experiment\thanks{This is the author's version. The published version is available at the EPJ Data Science journal in \url{https://doi.org/10.1140/epjds/s13688-017-0119-3}.}}

\author{\large Eduardo Graells-Garrido\thanks{Corresponding author: \url{egraells@udd.cl}.}, Leo Ferres, Diego Caro, Loreto Bravo}
\affil{\footnotesize Data Science Institute, Faculty of Engineering, Universidad del Desarrollo \\ Telef\'onica R\&D Chile}

\begin{document}

\maketitle

\begin{abstract}
Pokémon Go, a location-based game that uses augmented reality techniques, received unprecedented media coverage due to claims that it allowed for greater access to public spaces, increasing the number of people out on the streets, and generally improving health, social, and security indices. However, the true impact of Pokémon Go on people's mobility patterns in a city is still largely unknown.
In this paper, we perform a natural experiment using data from mobile phone networks to evaluate the effect of Pokémon Go on the pulse of a big city: Santiago, capital of Chile.
We found significant effects of the game on the floating population of Santiago compared to movement prior to the game's release in August 2016: in the following week, up to 13.8\% more people spent time outside at certain times of the day, even if they do not seem to go out of their usual way. These effects were found by performing regressions using count models over the states of the cellphone network during each day under study.
The models used controlled for land use, daily patterns, and points of interest in the city.

Our results indicate that, on business days, there are more people on the street at commuting times, meaning that people did not change their daily routines but slightly adapted them to play the game.  Conversely, on Saturday and Sunday night, people indeed went out to play, but favored places close to where they live.

Even if the statistical effects of the game do not reflect the massive change in mobility behavior portrayed by the media, at least in terms of expanse, they do show how ``the street'' may become a new place of leisure. This change should have an impact on long-term infrastructure investment by city officials, and on the drafting of public policies aimed at  stimulating pedestrian traffic.

\spara{Keywords} Pokémon, Mobile Phone Data, Call Detail Records, Floating Population, Urban Informatics.
\end{abstract}
\clearpage
\section{Introduction}

Pokémon Go is a location-based mobile game about capturing and ``evolving'' virtual characters that appear to exists in the same real-world location as players. As soon as it came out, people of all ages seemed to be caught in the frenzy of walking everywhere trying to find the next pocket monster. As it became world-wide hit, the game has fueled a flurry of speculation about its potential effects. It has been suggested, for example, that the game made kids and adults move out of the living room and into the open air, and that touristic attractions would attract more people if they had a Pokéstop (a place to check-in and get items), among many others. Real or not, Pokémon go has had some real-world effects, some of which are not ideal: governments issuing alerts on playing the game in minefields,\footnote{\url{http://goo.gl/AvjpGL}} searching for Pokémon in ``inappropriate'' places like the Holocaust Museums and the White House,\footnote{\url{http://goo.gl/TWeI86}} and even to causing accidents.\footnote{\url{http://goo.gl/d69rjS}} The game was so popular, that in some countries it reached engagement rates that surpass those of mainstream social platforms like Twitter and Facebook.\footnote{\url{https://goo.gl/lqwfUb}}

The general perception is that games like Pokémon Go and its predecessor, Ingress~\cite{majorek2015ingress}, could make whole populations change their mobility patterns through a reward-system: earning more points by catching creatures, getting to certain places and checking in, among other well-known gamification techniques.
Ingress and Pokémon Go have a laxer definition of ``check-points'' than a city's usual Points of Interest (POIs, such as museums and parks) including, for example, graffiti \cite{moore2015painting} and hidden heritage \cite{stark2016playful}. Thus, these games may help motivate visiting different kinds of places from the usual POIs.
Because people tend to visit few POIs in their daily routines~\cite{papandrea2013many}, playing the game implies that people would tend to visit different places from those they would usually visit. If this is the case, and considering that Pokémon is one of the most successful media franchises in the world \cite{buckingham2004pikachu}, providing empirical evidence in favor (or against) this folk hypothesis would help understand the level to which these games make people change their habits.

In this study, we seek to quantify the \emph{Pokémon Go Effect} on the pulse of a city, as seen from its floating population patterns. The ``floating population'' concept denotes the number of people present in a given area during a specific period of time, but who do not necessarily reside there. For instance, people who work in a business district are part of its floating population, since they probably reside elsewhere. Given the successful launch of Pokémon Go in Chile and the availability of mobile phone network data, we are able to ask sophisticated questions about floating population such as whether Pokémon Go has an effect on it at the city-scale; and if so, what the characteristics of these effects are.

In Chile, Pokémon Go was officially launched on August 3rd, 2016. Reports indicate that more than one million people downloaded the game within the next five days.\footnote{\url{https://goo.gl/T2pthM}}
To understand the game effects, we based our analyses on a set of mobile communications records from Telef\'onica Movistar, the largest telecommunications company in Chile, with a market share of 33\% in 2016. We used a dataset that follows the Call Detail Records structure, \ie, a dataset built for billing purposes, from July 27th, until August 10th. CDR datasets usually include logs of mobile phone calls, SMS's, and data-type network events (\eg, Web browsing, application usage, etc.), aggregated by a context-dependent amount of downloaded information~\cite{calabrese2015urban}.
Even though we cannot know who is playing the game, we hypothesize that this is not needed to evaluate the city-level effect: our aim is to measure how many people were on the streets before and after the release of the game.

We followed a natural experiment approach whereby we evaluated floating population patterns at two specific intervals of time: seven days before and seven days after the launch of Pokémon Go. We selected a specific number of devices to ensure that we only analyzed floating population patterns of active users who live in the city (see Section \ref{sec:dataset} for the filtering process). Given this, we assumed, conservatively, that a higher number of connected mobile devices meant a higher number of people on the streets. Then, using regression models suitable for {\em count} data, we evaluated whether the launch of the game had an impact on the floating population by individually testing each regression factor for significance. Our aim was to measure the factor associated to the game itself \cite{greene2008functional}.

We explored how spatio-temporal properties of mobility were related to the Pokémon Go effect. To do so, we also used complementary datasets that are either generally available, such as travel surveys, or that could be approximated using mobile phone data (\eg, land use). Hence, the methods presented in this paper can be used to perform a similar analysis in other cities, as well as monitoring for pattern changes.

The contributions of this work are two-fold. First, we introduce a methodology to analyze mobile records that allowed us to identify behavioral changes at the city level. Second, we provide a case study of a city in a developing country: Santiago of Chile, and report several empirical insights about the observed phenomena. To the extent of our knowledge, this is the first large-scale study on the effect of an external factor (a location-based augmented reality game) on the floating population of a city.

The main findings of our work are as follows: first, there is a significant effect of the availability of Pokémon Go on Santiago's floating population patterns, including covariates that account for daily patterns, land use, and available points of interest. The highest effect during business hours was found at 12:31, with 13.8\% more people connected to mobile networks. After business hours, the strongest effect was found at 21:31, with 9.6\% more people connected to mobile networks. Second, during business hours, the effect is significant at commuting times between important places (such as home and work) and break hours (\eg, lunch times).
We conclude that people adapted their routines to play the game, concentrating geographically in places with a high floating population. Unlike those effects found during business hours, Pokémon Go players were scattered around the city at night, which hints at the possibility that people played the game near their places of residence at times when they were usually indoors. Finally, we discuss our findings in the light of practical and theoretical implications in the areas of urbanism and the social life of the city.

\section{Data and Methods}

We focused our study on Santiago, the most populated city in the country, with almost 8 million inhabitants. Comprising an area of 867.75 square kilometers, urban Santiago is composed of 35 independent administrative units called municipalities. The city has experienced accelerated growth in the last few decades, a trend that has been predicted to continue at least until 2045 \cite{puertas2014assessing}.
Chile, and Santiago in particular, is one of the developing regions of South America with the highest mobile phone penetration index. There are about 132 mobile subscriptions per 100 people.\footnote{\url{https://goo.gl/sjWEjS}}
Santiago's growth and the general availability of mobile phones makes it an excellent city to perform research based on mobile communication data.

\subsection{Datasets}\label{sec:dataset}

We use the following complementary datasets:

\spara{Santiago Travel Survey and Traffic Analysis Zones}
The Santiago 2012 Travel Survey\footnote{\url{https://goo.gl/vStth8}.} (also known as Origin-Destination survey, or ODS) contains 96,013 trips from 40,889 users.
The results of this survey are used in the design of public policies related to transportation and land use. The survey includes traffic analysis zones of the entire Metropolitan Region, encompassing Santiago and nearby cities. We use this zoning resource for two reasons. On the one hand, the extent and boundaries of each area within a zone take residential and floating population density, administrative boundaries, and city infrastructure into account. This enables the comparison of several phenomena between zones. On the other hand, it allows us to integrate other sources of information, providing results that can be compared to other datasets such as land use properties \cite{graells2016sensing}.

\begin{figure}
\centering
\includegraphics[width=0.75\textwidth]{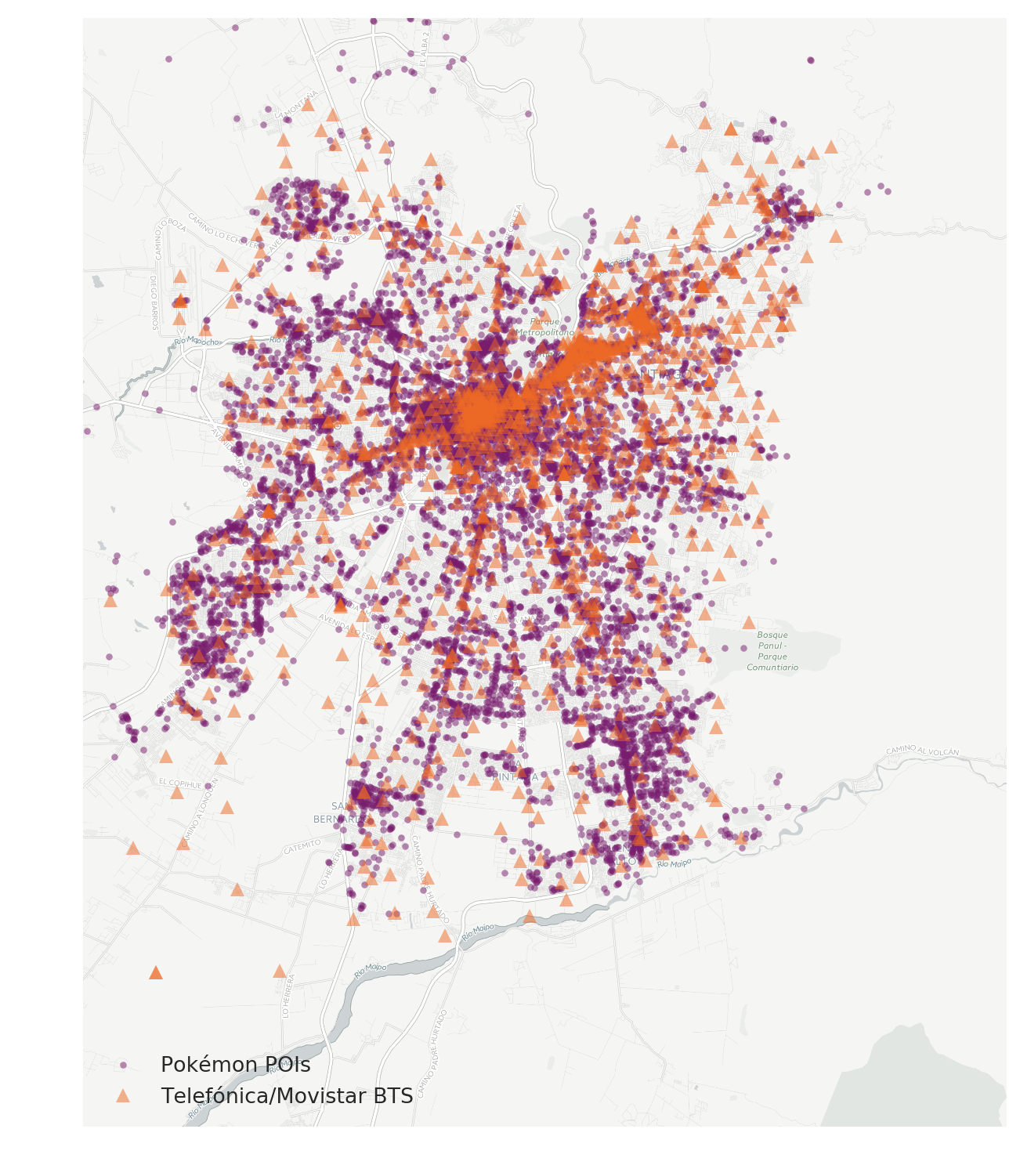}%
\caption{Spatial distribution of the Telef\'onica cell towers (orange triangles) and Pokémon points of interests (purple dots) in Santiago.
}%
\label{fig:maps}%
\end{figure}

The complete survey includes 866 zones; however, we were interested in urban areas of a single city. Since these are densely populated, we restricted our analysis to zones with a surface of under 20 square Km. As result, the maximum zone area is 18.37 square Km., with mean 1.34 and median 0.72 square Km.
Finally, we were interested in zones that have both cell phone towers and Pokémon points of interests (see Figure \ref{fig:maps}), resulting in 499 zones covering 667 km$^2$, about 77\% of Santiago.

\spara{Ingress Portals/Pokémon Points of Interest}
Before Pokémon Go, Niantic Labs launched Ingress \cite{majorek2015ingress} in 2012. Ingress is a location-based game where players choose one team (from two available), and try to \emph{hack} (take control) several \emph{portals} placed in real locations world-wide. Portal locations are crowd-sourced and include ``a location with a cool story, a place of historical or educational value,'' ``a cool piece of art or unique architecture,'' ``a hidden-gem or a hyper-local spot,'' among others.\footnote{\url{https://goo.gl/4QHtBQ}}
The definition of a portal, thus, includes points of interest that go beyond the definition used in, for instance, check-in based social networks \cite{moore2015painting}.
Once a portal is ``hacked,'' it belongs to the corresponding faction. A set of portals belonging to the same faction defines the limit of an area controlled by it. Since players need to be close to portals to hack them, this makes players explore the city to find portals to hack and conquer for their own teams.

Pokémon Go shares many game mechanics with Ingress, including the team concept. The main difference between the games is that while in Ingress players capture portals, in Pokémon Go they capture wild pocket monsters.
A subset of Ingress portals is defined to be a PokéStop (a place to check-in and get items) or a PokéGym (a place to battle against the Pokémon of other factions). In this paper, both are referred to as PokéPoints. Additionally, there are hidden Pokémon respawn points, where different creatures tend to appear. The main mechanic of the game is that players must walk around and explore to find creatures to capture. Note that all players see the same creatures, and one creature may be captured by many players. The game motivates walking in two ways: first, walking to points of interests that are scattered around city (see Figure \ref{fig:maps}); and second, by walking 1, 5 or 10 Kilometers, players can also hatch eggs containing random pocket monsters that have better biological properties than those caught on the wild.

\spara{Mobile Communication Records}
Telef\'onica has 1,464 cell phone towers in the municipalities under consideration. We studied an anonymized Call Detail Records (CDR) dataset from Telef\'onica Chile. This dataset contains records from seven days prior to the launch of Pokémon Go in 2016 (from July 27th to August 2nd) and seven days after (from August 4th to August 10th). We did not take into account the day of the official launch of Pokémon Go, as there was no specific hour in which the game was officially and generally available. Also note that the dataset contains pre-paid and contract subscriptions from Telef\'onica.

The dataset contains data-type events rather than voice CDRs \cite{calabrese2015urban}. Unlike typical Call Detail Records for voice, each data event has only one assigned tower, as there is no need for a destination tower.
Each event has a size attribute that indicates the number of KiB downloaded since the last registered event.

We did not analyze the records from the entire customer population in Santiago. Instead, we applied the following filtering procedure: (i) we filtered out those records that do not fall within the limits of the zones from the travel survey, and also those with a timestamp outside the range between 6:00 AM and 11:59 PM;
(ii)~to be considered, mobile devices must have been active every day under study, because a device that does not show regular events may belong to a tourist, someone who is not from the city, or does not evidence human-like mobility patterns such as points-of-sale (which are mostly static);
(iii) only devices that downloaded more than 2.5 MiB and less than 500 MiB per day were included, as that indicates either inactivity or an unusual activity for a human (\ie, the device could be running an automated process);
(iv) we used a Telef\'onica categorization scheme that associates an anonymized device ID with a certain category of service: for example pay-as-you-go, contractual, enterprise, etc. This gives us a good idea of the general kind of account holders. Thus, every step of this procedure was taken to ensure that events were triggered by humans. After applying these filters, the dataset comprised records from 142,988 devices.

Our filtering procedure ensures that a positive difference in the number of connections between two different days represents more people within a given zone. Depending on conditions such as time and location, we may interpret that some of those people are on the street. For instance, we may look at typical times where people commute, or at places where people are either inevitably outdoors (\eg, in a park) or inevitably at residential areas, which tend to have WiFi networks.\footnote{72\% of homes have Internet access according to the last survey of telecommunication infrastructure by the Ministry of Telecommunications of Chile, available at \url{https://goo.gl/bwyzPG}.}

\spara{Land Use Clusters}
We may take each traffic analysis zone to belong to one category of land use: residential, business, and areas with mixed activities (\eg, business plus recreation or shopping activities, etc.). These categories are the result of our previous work on land use and CDR data, which is based on hierarchical clustering of time-series of connections at each zone of the city~\cite{graells2016sensing}.

\subsection{Approach}
This study uses a natural experiment approach to measure the Pokémon Go effect at the city scale. To do this, we analyzed the change in population patterns before and after the launch of Pokémon Go as evidenced by CDR data.
First, we described a method of smoothing the number of connected devices at each cell tower according to several  snapshots of the tower network. A snapshot is the status of the cell phone network in a given time-window \cite{naboulsi2014classifying}. Then, we aggregated these device counts at the zone level to define a set of observations that we evaluated in a regression model. We took into account covariates that enabled us to isolate and quantify the Pokémon Go effect.

\spara{Device counts at each tower and zone level aggregation}
Let $e\in E$ be a network event, and $|E|$ is the total number of such events. A network event $e$ is a tuple $(d,u,b,z)$, where $d$ is a timestamp with a granularity of one minute, $u$ is some (anonymized) user id, $b$ is a tower id, and $z$ is one of the previously defined geographical areas of Santiago. For each tower $b$ and time $d$ we developed a time-series $B_{d,b}$ which represented the number of unique users from $E$ connected to $b$ at $d$.
Due to the sparsity of CDR data, it is possible that $B$ is not continuous. As consequence, the time-series could be null ($B = 0$) at a point of time where there were active devices at the corresponding tower. To account for this sparsity and obtain a continuous time-series, we smoothed each time-series $B$ using Locally Weighted Scatterplot Smoothing (LOWESS) interpolation~\cite{cleveland1988locally}, obtaining $B'_{d,b}$. To obtain a LOWESS curve, several non-parametric polynomial regressions are performed in a moving window. The size of this window is the bandwith parameter for the model. In our implementation, this value is 30, which is interpreted as follows: each connection influences (\ie, is counted into) its correspondent location during 30 minutes. Then, for each zone $z$, we aggregated all time-series $B'_{d,b_i}$ into $S_{d,z}$ by computing the sum of all time-series $B'_{d,b_i}$, where the tower $b_i$ lies in $z$ (determined using a point-in-polygon test, as in~\cite{mao2015quantifying}). Finally, each $S_{d,z}$ time-series represents the floating population profile for each zone under study.

\spara{Measuring the Pokémon Go effect}
To measure the city-wide Pokémon Go effect we considered the availability of the game as a city intervention, which started on the day of the launch of the game. Our hypothesis is that the following days would show an effect of the game if people went out, regardless of being players or not, and this effect would show on the number of people connected at each zone of the city. To do so,
we used Negative Binomial Regression~(NB)~\cite{nelder1972generalized,greene2008functional} applied to our dataset at 1-minute intervals during a day. The NB regression model has been used frequently to analyze over-dispersed count data, \ie, when the variance is much larger than the mean, contrary to the Poisson model \cite{cameron2013regression}.

For every minute under study (note that we restricted ourselves from 6 AM until 12AM), we performed a NB regression using the observed device counts at each zone of the city in all available days in the dataset.
The model is specified as follows:
$$
\log E[X(t)] = \log a + \beta_0 + \beta_1 PoGo + \beta_2 DayOfWeek + \beta_3 LandUse + \beta_4 PokePoints,
$$
\noindent where $E[X(t)]$ is the expected value of the number of active devices within a zone at time $t$. The \texttt{PoGo} factor is a binary variable that has a value of 0 when the game was not available, and 1 when it was available. The covariates \texttt{DayOfWeek} (with values business\_day, Saturday, and Sunday) and \texttt{LandUse} (with values residential, business\_only, and  mixed\_activities) account for the fluctuations in population on different days according to land use. Both factors use dummy coding because they are categorical. The covariate \texttt{PokéPoints} represents the number of PokéStops and PokéGyms, which are proxies for points of interest within an area, accounting for the number of potential attracting places in each zone.
The exposure value $a$ represents the surface area of each zone. Because urbanists designed each zone having into account population density, transportation infrastructure, and administrative boundaries, the exposure parameter also allowed to control indirectly for these potential covariates.

The model output allows the following interpretation: the $\beta$ coefficient assigned to each factor represents the difference of the logarithm of expected counts in a zone at time $t$, if all other factors were held equal. Since $\beta = \log{\mu_1} - \log{\mu_0} = \log{\frac{\mu_1}{\mu_0}}$, then the difference of logarithms equals the logarithm of the ratio between population counts before and after the availability of the game. The exponential of this coefficient is defined as Incidence Rate Ratio, $IRR_\beta(t) = e^{\beta(t)}$. We developed a time-series of $IRR_\beta(t)$ values for each factor. By analyzing these time-series we determined when, in terms of time-windows within a day, there were significant effects for each factor.

\section{Case Study: The Effect of Pok\'emon Go in Santiago, Chile}

Our aim was to measure the effect of Pokémon Go in the number of people and their mobility patterns in a city. As stated in our methods section, the first step is to obtain a smoothed number of connected devices to each tower per minute during the previous/following seven days to the launch of the game.

\begin{figure}
\centering
\includegraphics[width=0.75\textwidth]{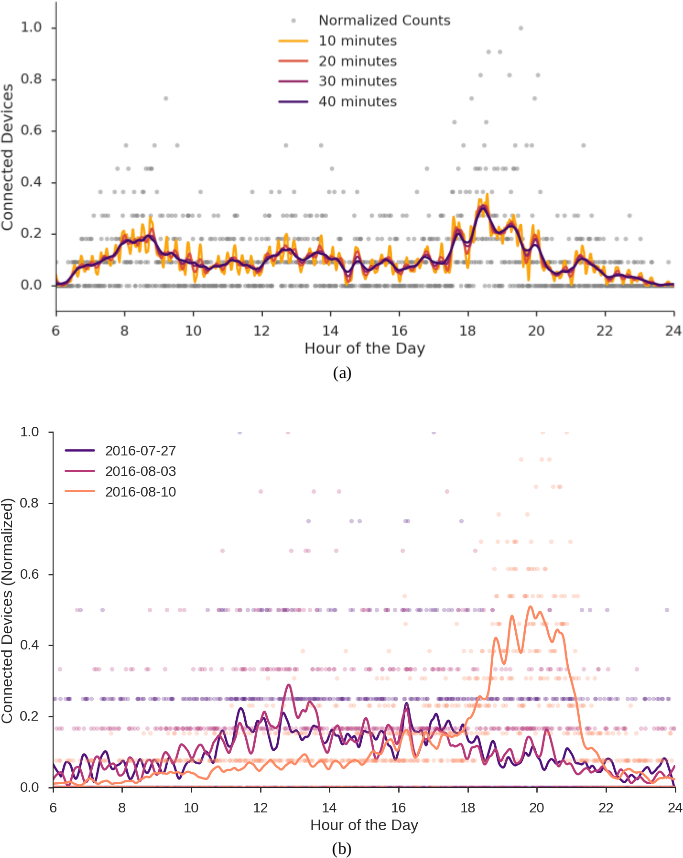}
\caption{Connected device counts at a specific cell tower (a), and at a zone, aggregating several towers (b). Each dot represents the number of connections at a specific time of the day. The curves represent LOWESS interpolation of the number of connected devices.}%
\label{fig:bts_counts}%
\end{figure}

Figure \ref{fig:bts_counts} (a) shows the normalized number of events per minute in a specific tower (Moneda Metro Station) during July 27th.
The chart shows that there are many minutes (dots) without registered events, \ie, their position in the $y$-axis is 0. This shows the sparsity of the data, because this tower is located in a business area with a high rate of public transportation traffic. We used LOWESS interpolation to be able to approximate the number of people even in those minutes where there are no events. The continuous lines in Figure \ref{fig:bts_counts} (a) are LOWESS interpolation calculations considering values of 10, 20, 30 and 40 minutes for the LOWESS time windows.
After manual inspection of the dataset, we reasoned that a time window of 30 minutes captured interesting regularities without incurring in noise produced by the sparsity of the data, nor smoothing the dataset too much.

Figure \ref{fig:bts_counts} (b) shows the aggregation of towers within Zone 44 for three different days: a week before the launch of the game, the day of the launch, and a week after. Zone 44 contains a park (Parque O'Higgins), which is not usually visited at night, hinting that the availability of the game may have influenced a different behavior.

\begin{figure}[tp]
    \centering
    \includegraphics[width=\textwidth]{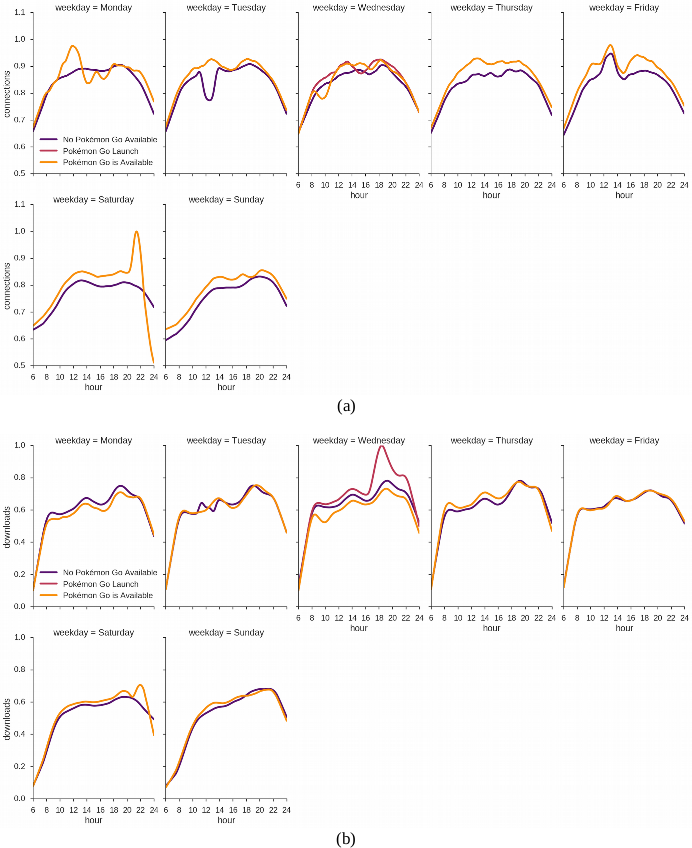}
    \caption{LOWESS time-series of normalized number of connected devices (a) and normalized total downloaded information (b), per minute, for all days in the dataset. The distributions are normalized by dividing the actual number of connections by the global maximum value. Each panel represents a different nominal day, with curves for a day before/after the launch of Pokémon Go. Wednesdays include an additional day that represents the launch of the game.}
    \label{fig:data_connections_and_consumption}
\end{figure}

\spara{City-level connections}
Figure \ref{fig:data_connections_and_consumption} shows the city-level aggregated distributions of: (a) number of connected devices, and (b) downloaded information. Note that in both figures the distributions are normalized by dividing each value by the global maximum value.
The figure considers three categories of days: before, during, and after the launch of Pokémon Go.
In terms of connections, one can see that the patterns are stable across most days, with the total number of connections generally higher when Pokémon Go was available.
This means, intuitively, that there were more people connected to the network presumably playing the game. Two rather surprising effects are found on Mondays between 10 AM and 12 PM when Pokémon Go was available, and Tuesdays at about 12 PM when it was not yet available. In the first case, we hypothesize that since it was the first Monday after the launch of the game, people were trying it out. In the second case, there does not seem to be any explanation for the sudden drop of connections before the launch of the game, but it might be due to general network outages. Notice that the curve of Pokémon Go availability in the same time period behaves as expected.

In terms of downloaded information, the patterns present smaller variations across days, and some of them present identical behavior. We cannot assume that more data implies people playing, due to how the game works: the user has to keep the screen active for the game, and most of the game data-transfers are text-encoded packages of information (\eg, position updates about nearby pocket monsters, in JSON format). Conversely, social media applications can use more data, due to the size of images, and the streaming of audio and video.
Hence, we leave data-traffic analysis for future work.
However, one effect that is clear is how the game was massively downloaded on its launch, Wednesday, August 3rd.

Note that a word should be said about Saturdays. The first Saturday after the release of Pokémon Go exposes the highest number of device connections found in the dataset.
While it is tempting to associate any found effect to other extraneous factors (for example, the Olympic Games were on TV at the same time of our study), Figure \ref{fig:data_connections_and_consumption} (b) shows that Saturdays exhibit similar behavior in terms of data consumption. Thus, it is unlikely that the Olympic Games were watched massively, at least not using mobile phones.

\begin{figure}[tp]
    \centering
    \includegraphics[width=\textwidth]{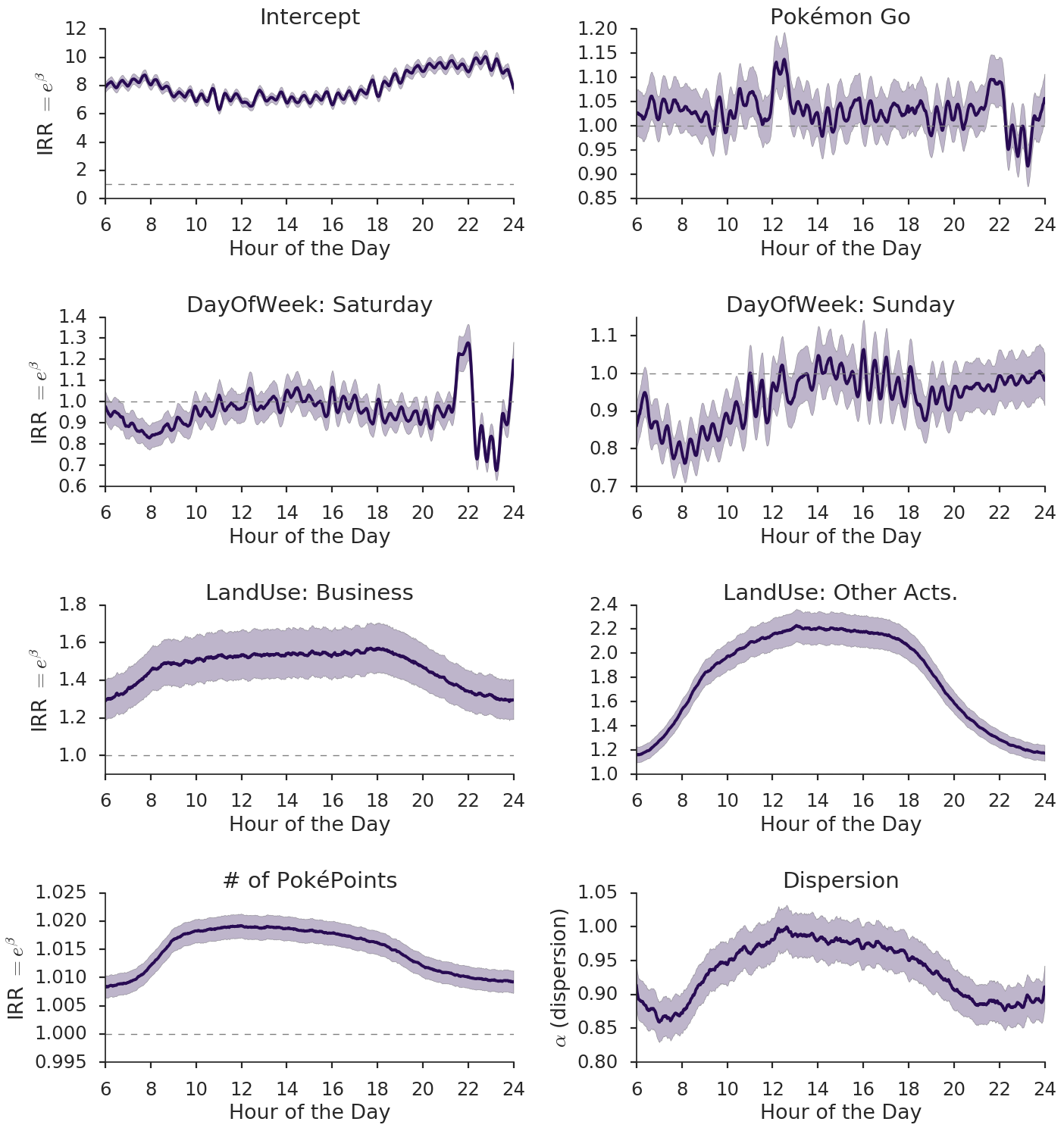}
    \caption{Covariates and dispersion ($\alpha$) for each factor of the Negative Binomial regression applied to the 1-minute snapshots. The Intercept factor includes the base values of each factor: 0 for the Pokémon Go factor, business\_day for \texttt{DayOfWeek}, residential\_area for \texttt{LandUse}, and 0 for the number of \texttt{PokéPoints}.}
    \label{fig:pokemon_regression_factors}
\end{figure}

\spara{Negative Binomial Regressions}
After aggregating the smoothed counts for each zone, we performed a NB regression for every 1-minute snapshot across the days in our dataset. In other words, for each minute, the observations are the aggregated zone counts for all days, for all zones, at that specific minute.
As result, we obtained a time-series of regressions representing the dynamics of each factor in our model.
Figure \ref{fig:pokemon_regression_factors} shows the Incidence Rate Ratios (IRR) for each factor, as well as the distribution of model dispersion ($\alpha$).
An IRR of one means that the factor does not explain change. Although the significance of each factor is evaluated minute by minute applying a $z$-test, the visualization of each factor's IRR allows to visually determine significance if its 95\% confidence intervals does not intercept $y = 1$.

The day of week covariates captured behavior expected for weekends. There are fewer people (IRR is significantly lesser than 1) in the morning for both kinds of days. For instance, the Saturday factor captured a portion of the night effect (its IRR at 9:59 PM is 1.278). This means that not all people who went out on the Saturday when the game was available can be explained by the Pokémon Go factor.
The land use covariates captured the dynamics of high floating population during the day, given the bell-shape of their distributions, and their IRRs are significant during the whole day.
The PokéPoints covariate, which is a proxy for points of interest in the city, is also significant during the whole day. Its maximum IRR is 1.019, which means that for every additional POI in a zone, the number of people increases by 1.9\%, if all other factors are held equal when performing the comparison.

We then analyzed the time windows when the Pokémon effect is significant. Table \ref{table:pogo_effect} summarizes these time windows, most of which are a few minutes long. However, there are two windows with prominent lengths: from 11:58 AM to 12:46 PM, and from 9:24 PM to 10:12 PM. These time windows also contain the highest Pokémon Go IRR values found per window: 1.138 and 1.096, respectively. Thus, all other factors being held equal, the availability of the game increased the number of people connected to mobile towers in the city by 13.8\% at lunch time and 9.6\% at night.

Finally, notice that we also tested for statistical interactions between the regression factors, but they were not significant. Additionally, we tested the model without the covariates, having only the intercept and the Pokémon effect. Particularly, the greater time-windows presented similar results and lengths, indicating that the model is robust.

\begin{table}[tp]
\centering
\footnotesize
\caption{Time-windows where the Pokémon Go effect was significant. Each window includes its maximum Incidence Rate Ratio (IRR) and its corresponding time of the day. }
\label{table:pogo_effect}
\begin{tabular}{lcc}
\hline
Time Window & Max IRR & Time of Max IRR \\
\hline
 6:34 -- 6:47 & 1.062  &  6:40 \\
 7:07 -- 7:18 & 1.056  &  7:11 \\
 7:37 -- 7:46 & 1.054  &  7:42 \\
 7:48 -- 7:48 & 1.047  &  7:48 \\
 9:35 -- 9:43 & 1.060  &  9:40 \\
10:27 -- 10:41 & 1.077  &  10:34 \\
10:53 -- 11:18 & 1.071  &  11:07 \\
11:58 -- 12:46 & 1.138  &  12:31 \\
13:06 -- 13:09 & 1.051  &  13:08 \\
15:36 -- 15:51 & 1.058  &  15:50 \\
16:17 -- 16:21 & 1.052  &  16:19 \\
18:30 -- 18:34 & 1.052  &  18:31 \\
19:42 -- 19:45 & 1.051  &  19:43 \\
21:24 -- 22:12 & 1.096  &  21:38 \\
22:22 -- 22:25 & 0.955  &  22:25 \\
22:44 -- 22:52 & 0.954  &  22:52 \\
23:09 -- 23:21 & 0.954  &  23:09 \\
23:57 -- 23:59 & 1.057  &  23:59 \\
\hline
\end{tabular}
\end{table}

\spara{Explaining the Pokémon Go Effect}
Given the specification of our model, our results are city-wide. To explore results at a finer geographical level, we estimated the difference between two time-series: the mean of connection counts per zone after and before the launch of the game by separating observations between business days and weekends. Then, we adjusted the time-series according to the surface area of each zone, and we obtained time-series per zone that indicate whether they had, on average, more or fewer people connected after the launch of the game.

\begin{figure}[tp]
    \centering
    \includegraphics[width=0.65\textwidth]{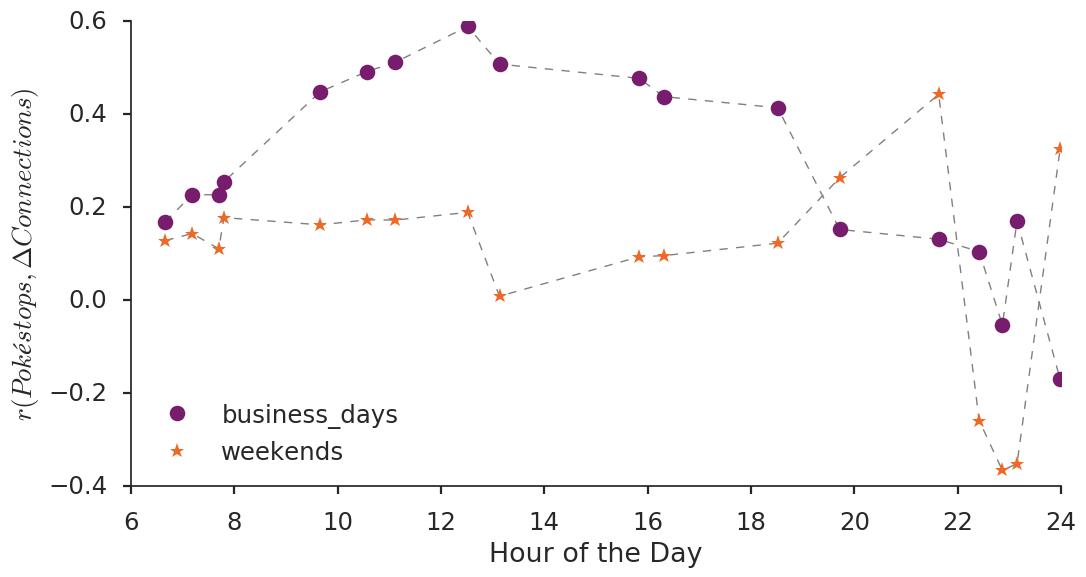}
    \caption{Pearson correlation coefficients for business days and weekends. The correlation estimates the relation between the number of PokéPoints in a given area, and the number of connections within it. These coefficients are proxies for how much attraction was generated from areas with many points of interest.}
    \label{fig:pokemon_correlation}
\end{figure}

To find whether these differences correlate with the number of PokéPoints per square kilometers in each zone, we performed a Pearson correlation for all the minutes of the day with maximum IRR values (see Table \ref{table:pogo_effect}). These correlations vary during the day, as shown on Figure \ref{fig:pokemon_correlation}. For business days, the highest correlation was found at 12:31 ($r = 0.59$, p $< 0.001$). For weekends, the highest correlation was found at 21:38 ($r = 0.44$, p $< 0.001$). The effect is stronger at lunch time on business days, and at night on weekends.

\begin{figure}
\centering
\includegraphics[width=\textwidth]{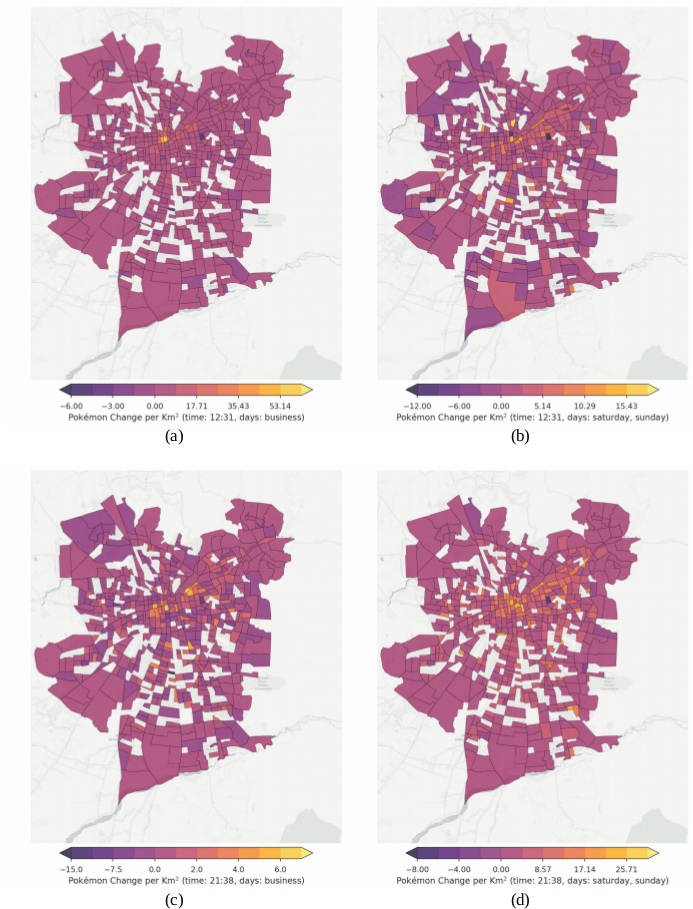}
\caption{Choropleth maps of the difference between the means of connected devices per zone, before and after the launch of of Pokémon Go. Top row shows differences at 12:31 PM on business days (a) and weekends (b). Bottom row shows differences at 9:38 PM on business days (c) and weekends (d). The differences are adjusted by zone area, and consider business and weekend days as different groups. A zone with a value of 0 did not show differences between periods.}%
\label{fig:result_maps}%
\end{figure}

Figure \ref{fig:result_maps} displays four choropleth maps of Santiago. The top row contains two maps, both showcasing differences per zone at 12:31 PM. The left map (a) displays business days, and the right (b) displays weekends. Similarly, the bottom row displays differences at 21:38 (business days at (c), weekends at (d)).
Regarding the correlations described in Figure \ref{fig:pokemon_correlation}, one can see that Fig. \ref{fig:result_maps} (a) shows a highly concentrated effect in the city historical center. Reportedly, players visited this place at all times of the day, both because of its location within the city, as well as the availability of PokéPoints.\footnote{\url{https://goo.gl/RcxkVA}}
In contrast, weekends present a more diversified effect on the city, specially at night, with many areas showing highly positive differences. A careful exploration of the map (d) reveals that zones with higher differences contain or are near parks and public plazas.

\begin{figure}[tp]
    \centering
    \includegraphics[width=0.95\textwidth]{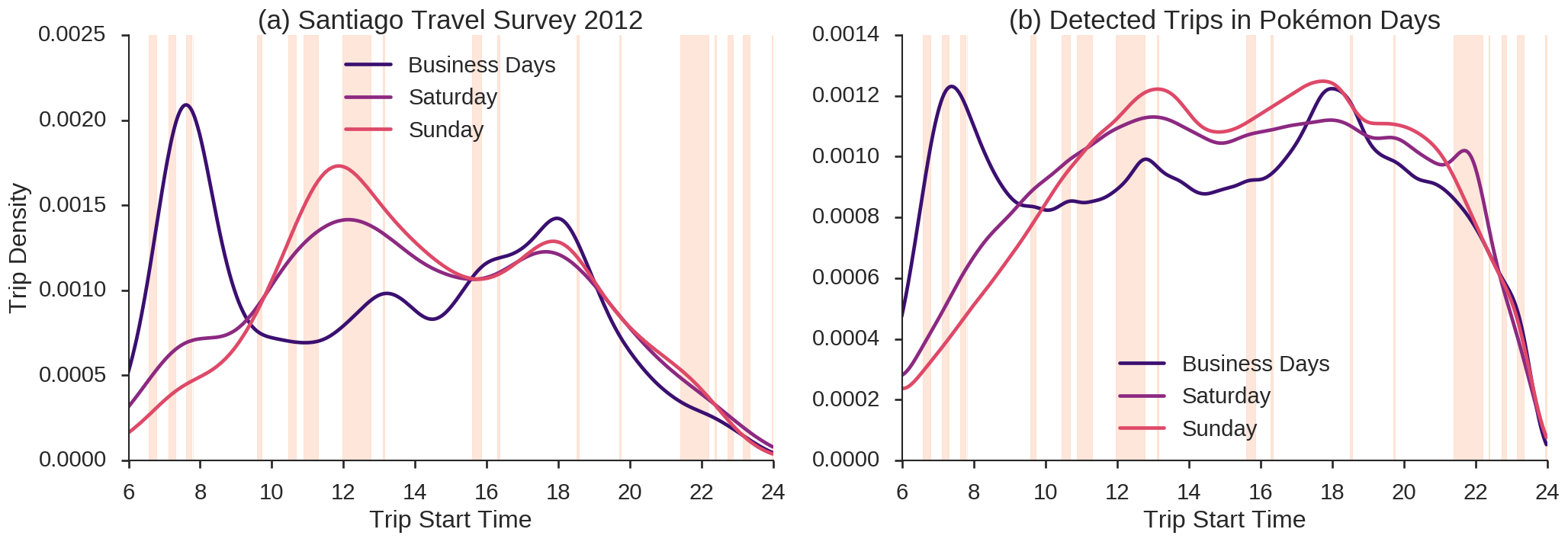}
    \caption{Distribution of trip start time according to: (a) the travel survey held in Santiago in 2012, and (b) detected trips during the Pok\'emon days. The trips are aggregated into business days, Saturdays, and Sundays. Each rectangle shows the time-window in which the Pokémon Go effect is significant in the regression models.}
    \label{fig:pokemon_irr_trip_time}
\end{figure}

We also compared how the Pokémon Go effect related to urban mobility,
by graphically comparing the significant time-windows with the trip
density distribution from the travel survey. Figure
\ref{fig:pokemon_irr_trip_time}A shows the position of each
time-window as a bar, and the trip densities as lines for business days,
Saturdays, and Sundays. In the morning, the significant
time-windows appear when the trip density is increasing or when it
reaches a local maxima in business days.  They also appear after noon,
potentially linked to lunch breaks during business hours.  All other
effects, such as those at night, do not appear to have any correlation
with trip behavior. Given that the travel survey is already about five
years old, it was also important to compare our results to {\em
  current} trip distributions. We applied a work-in-progress algorithm, based on previous work on inferring trips from CDR data~\cite{Graells-Garrido:2016:DYD:2962735.2962737}. Figure
\ref{fig:pokemon_irr_trip_time}B shows that, in general terms, trip
distributions and the effects still hold for CDR detected trips. This
is not surprising, given how hard it is to change a whole population's
general mobility patterns without any exogenous circumstances, at
least at the 5-year scale. In any case, this lends further validation
to our main results.

\section{Discussion}
We performed a natural experiment at the city-scale comparing the behavior of a subset of the population before and after the launch of Pokémon Go in Santiago. We found that the availability of the game increased the number of people that connected to the Internet on their mobile phones by 13.8\% at lunch time and 9.6\% at night.
A further exploration of the relationships among urban mobility patterns, mobile connectivity and points of interests revealed that there are two primary ways in which the Pokémon effect is noticeable. On the one hand, people take advantage of commuting times and breaks during the day to play the game. Thus, players tend to be near their work/study places, which are mostly downtown. On the other hand, on weekends at night the effect is more diversified, implying that people tend to play the game in places near their homes.

In his book ``The Great Good Place'', Ray Oldenburg discussed the need for \emph{third places} in the city: \emph{``In order for the city and its neighborhoods to offer the rich and varied association that is their promise and potential, there must be neutral ground upon which people may gather. There must be places where individuals may come and go as they please,
in which no one is required to play host, and in which we all feel at home and comfortable''}~\cite{Oldenburg1989}. The concept comes from the designation of home and work (or study) as first and second places in one's life. Nowadays, third places are facing two challenges. First, virtual worlds~\cite{steinkuehler2006everybody} and social networks~\cite{rao2008facebook} provide social infrastructure that is similar to that of third places, but without going out of the first or second place. Second, perceptions of crime and violence are making the city feel less safe than it really is. In Santiago, this has been called \emph{fear of life} \cite{dammert2003fear}.
Hence, the usage of location-based augmented reality games may help to alleviate both situations, by placing virtual worlds on top of physical reality, and by motivating people to go out and walk around their neighborhoods as our results indicate.

Because Pokémon Go has the effect of increasing the number of pedestrians on the street, it has the potential to convert the city into a third place given its social features.
Motivating the presence of pedestrians is important and relates to a theory put forth by Jane Jacobs, who claims that there are four conditions that must be met for a city to be lively and safe. These conditions are: the presence of pedestrians at different times of the day, the availability of mixed uses in districts, the mixture of old and new buildings, and the availability of many crossings for pedestrians~\cite{jacobs1961death}. Even though testing theories like these is difficult, some work has already been done. For instance, the city of Seoul found some evidence in favor of the theory by using house-hold surveys \cite{sung2015operationalizing}. Recently, mobile communication records have been used to validate Jacobs' theories in Italian cities \cite{de2016death}.
Studies like those mentioned above~\cite{sung2015operationalizing,de2016death} perform an ex-post analysis of whether lively places comply with theories. A more granular approach would be to perform natural experiments like ours. To the extent of our knowledge, this is the first experiment of its kind performed using mobile records of data-type. Our method makes it possible to measure the effect not only of long-term but also short-term interventions, opening a path to quantify how much specific actions help improve quality of life in the city. For instance, crime-data could be correlated with results obtained with our method to test whether the safety and livelihood theories hold after specific interventions. Additionally, as our figures have shown (cf. Fig. \ref{fig:result_maps}), our method is able to produce maps apt for \emph{patch dynamics} visualization to monitor population density \cite{pulselli2008computing}.

\spara{Limitations}
One limitation in our study is the lack of application usage identification. Thus, even though we included several covariates in our model to account for other effects, we may still be confounding other effects outside the Pokémon Go factor.

Another possible confounding variable is the presence of events that
may attract lots of people at certain hours, influencing the Pokémon
effect. In particular, since popular events such as important football
games or cultural presentations were being held in Santiago at the
same time as Pokémon go was launched, we felt the need to test for an
effect. Adding those events as a new land use category at the zone and
the time when the popular events were held, we found that they only
account for the 0.01 percent of observations. Therefore, popular
events had either negligible effects or no effect at all in our
study.

While it can be argued that most of the Pokémon Go media appearances
focus on specific situations, like those related to museums, tourism,
and physical activity of some players, our empirical results uncovered
to which extent the availability of the game is related to the number of
people on the street. Even though mobile phone datasets are usually
not public, it is becoming more common to have access to this kind of
information thanks to several initiatives in opening and sharing data
\cite{calabrese2015urban}. This kind of analysis is almost costless to
perform by telecommunications companies, because mobile records are
already extracted and stored for billing purposes.

A further issue involves the nature of the dataset itself. As we
mentioned in the paper, Telefónica Movistar has a little more than a third of
the Chilean mobile marketshare. This is the largest portion of the
market (the other 60\% is owned by many other smaller companies, with
the second largest one at 20\%). This does, however, introduce some
data biases, although they are hard to identify and quantify. One such
bias is population bias: it is not the case that Telefónica customers
are distributed uniformly in the geographical space under study. Since
we do not have access to the Telefónica database of customers, we
cannot tell how representative a certain cell-phone tower is to Telefónica
versus other telcos.

A final note of caution could be said about weekend effects: since we
only have one weekend, we cannot be completely confident about the
results. Unfortunately, this is the dataset we have. In any case, our
study was meant to be at ``city'' level, and only analyzed the
different days to explore the nature of the effect.

Regarding a potential novelty effect in our results, it may be the case that not all Pokémon Go players during the first few days were still engaged with the game later. Indeed, the number of Pokémon Go players has gone down enormously. While the novelty effect might be true, our study was not aimed at evaluating the popularity of the game. Instead, its purpose was to quantify the city-level effects, something that could have been more concentrated in time: in Chile, the game was highly anticipated by the users and the media, because it was released almost one month later than in the USA.

\spara{Future Work} User modeling and classification may help to
categorize users into Pokémon players and non-players, in order to
study the individual effects of the game. This can be done, for
instance, by estimating their daily routines from their CDR-based
trajectories
\cite{calabrese2011estimating,Graells-Garrido:2016:DYD:2962735.2962737},
as well as their home and work locations
\cite{graells2016sensing}. Using these methodologies, we could learn
whether they visited unknown places, or whether they walked slowly or
faster.  An epidemiological analysis of player behavior
\cite{kulldorff1998statistical} could help evaluate whether social
interactions influence city exploration as a result of the Team Battle
Dynamics featured in the game.  A study of whether the use of
the game is correlated with crime-rate reduction in public places,
which could be linked with urban theories about street safety and
could be useful for urban planners or the police, as well as providing
evidence to some of Jacob's theories.

Finally, given the dataset we had available, we measured the immediate
change in mobility patterns for the whole city of Santiago. However,
it would have been very interesting to show that the changes in
behavioral patterns hold for some weeks further into the future,
perhaps even when the popularity of the game was declining (see the
provisos above). This could have been a very interesting result that
may have opened possible innovative usages of game-based strategies of
this kind to modify citizens' behavioral routines.

\section{Related Work}

\spara{Mobile Phone Data Analysis}
Our work builds on the extensive literature on the analysis of mobile records. We refer the reader to comprehensive surveys of research in this wide area \cite{blondel2015survey,barlacchi2015multi}. Here, we focus on describing the specific topics that intersect with our research. In order to analyze our data, we borrowed the concept of a ``snapshot,'' \ie, the status of the cell phone network during a specific interval of time \cite{naboulsi2014classifying}. The comparison of snapshots of the network enables us to find which time instants actually show interesting, and, in our case, statistically significant differences. In contrast, previous work has focused mostly on studying the network when it showed higher traffic volume or population density \cite{deville2014dynamic}, without controlling for covariates or population size, which we do in this work. Controlling for this was achieved by fixing the number of users to only those active {\em every single day} under study, minimizing ``noise'' in the form of one-off users, for instance. We also make use of the concept of {\em floating population profile} derived from our own,  and other similar work \cite{graells2016sensing,reades2009eigenplaces,soto2011automated,noulas2013exploiting,lenormand2015comparing,toole2012inferring}.
An interesting result is that research based on this methodology has proven consistent across different cities, enabling urban planners to compare cities with respect to their land use patterns \cite{lenormand2015comparing}, as well as to study how rhythms of life differ according to socio-cultural factors~\cite{ahas2010daily}.

According to a recent survey of urban sensing research \cite{calabrese2015urban}, local event analysis is a key area of mobile phone data analysis. Local events are usually defined as unusual gatherings or movements of massive amounts of people (\eg, protests, emergencies, sports, natural events, etc.) \cite{calabrese2010geography,ferrari2012people,traag2011social,bagrow2011collective}. Hence, the unit of analysis is a single event with time and space constraints.
Our method, instead, works at the city level with less prescribed time and location from those above. One thing in common between our work and the cited references is that all analysis have been performed ex-post, which makes it possible to create spatio-temporal signatures of places \cite{reades2009eigenplaces}. Even though these approaches allow us to analyze and understand the city, they do not allow measurement of city-scale phenomena due to their assumptions of locality.

Another relevant area is prediction and forecasting of human mobility~\cite{calabrese2010human,shimosaka2015forecasting,song2010limits,gonzalez08}, which allow to understand distributions~\cite{gonzalez08} and limits of predictability in human mobility~\cite{song2010limits}. Those methods focus on the big picture of mobility and rely on probabilistic models that need extensive datasets covering large periods of time, unlike ours, which only analyzed a two-week period.
Another approach uses regression~\cite{shimosaka2015forecasting}, as we do. However, our method differs: instead of using longitudinal data in only one regression model, for which Poisson models are better suited~\cite{shimosaka2015forecasting,cameron2013regression}, we use many consecutive Negative Binomial models, one for each time snapshot of the network, thus avoiding violating the assumptions of the Poisson model, and controlling for daily rhythms at the same time~\cite{sevtsuk2010does}. As the dispersion value showcases (cf. Figure \ref{fig:pokemon_regression_factors}), the NB regression was a correct choice, due to the exposure $\alpha$ being greater than zero~\cite{chan2009bayesian}. In our case, factors that could have caused over-dispersion include aggregation and non-uniform spatial distribution of the units of analysis~\cite{linden2011using}.

\spara{Augmented Reality and Location-based Games}
The effects of augmented reality games and applications within the city have been anticipated for more than one decade  \cite{flintham2003line,magerkurth2005pervasive}. However, the limits in mainstream hardware have impeded their general implementation/adoption at different times. Most studies about the impact of those games have been small-scale only \cite{kosoris2015study}. Even though smartphone technology has allowed location-based augmented reality games to become more commonplace in the last few years, until the launch of Pokémon Go, they still lacked the cultural impact needed to have a considerable effect on the city. As Frank Lantz is quoted in~\cite{apperley2013cybercafe}, in relation to the game PacManhattan: \emph{``If you want to make games like this you have to work hard to recruit an audience for them, you can’t just make up something awesome and then hope that people fall into it''} \cite{lantz2007pacmanhattan}. Since as we discussed above, Pokémon is one of the most successful media franchises in the world \cite{buckingham2004pikachu}, it enables the unique opportunity to study both, the impact of a location-based augmented reality game; and the effect of an intervention at the city scale when it comes to population mobility.

\section{Conclusions}

In this paper, we studied how Pokémon Go affected the floating
population patterns of a city. The game led to notable pedestrian
phenomena in many parts of the world. This is extraordinary in the
sense that it happened without the usual triggers like war, climate
change, famine, violence, or natural catastrophes.  In this regard, to
the extent of our knowledge, this is the first large-scale study on
the effect of augmented reality games on city-level urban mobility.
Given the massive popularity of the Pokémon brand, and its cultural
impact in many parts of the world, we believe that the found effects
represent a good approximation of mobility change in a large city.

Jane Jacobs theorized that the streets need more pedestrians to be
safe and lively~\cite{jacobs1961death}. Using CDR data, it has been
shown that this is true in at least some
cities~\cite{de2016death}. Thus, one of the most important conclusion
of our work is that cities may not need to change their infrastructure
in the short term to motivate pedestrians to go out. A game about
imaginary creatures lurking in neighborhoods, that can be collected
using cell phones encourage people to go out and make streets more
lively and safe when they are commuting, or when they have free
time during lunch or at night.

In summary, this study identified and investigated the effect of a
specific type of phenomenon in the pulse of a city, measured through
its floating population mobility patterns and usage of Pokémon Go. Our
methods can be used to perform other natural experiments related to
urban mobility, enabling measurement of the impact of city-wide
interventions, and using the results to inform public policy changes.


\section*{Funding}
The authors would like to thank Movistar - Telef\'onica Chile and the Chilean government initiative CORFO 13CEE2-21592 (2013-21592-1-INNOVA\_PRODUCCION2013-21592-1) for finantial support of this paper.

\section*{Competing interests}
The authors declare that they have no competing finantial interests.

\section*{Acknowledgements}
We thank Alonso Astroza for providing a crowdsourced list of Ingress Portals validated as Pokémon Go PokéStops and PokéGyms.
The analysis was performed using Jupyter Notebooks \cite{perez2007ipython}, jointly with the \emph{statsmodels} \cite{seabold2010statsmodels} and \emph{pandas} \cite{mckinney2010data} libraries. All the maps on this paper include data from \textcopyright OpenStreetMap contributors and tiles from \textcopyright CartoDB. We also thank Telef\'onica R\&D in Santiago for facilitating the data for this study, in particular Pablo García Briosso. Finally, we thank the anonymous reviewers for the insightful comments that helped to improve this paper.

\section*{Author Contributions}
EG and LF designed the experiment. EG, LF and DC performed data analysis.
All authors participated in manuscript preparation.

\section*{Availability of data and materials}
The Telef\'onica Movistar mobile phone records have been obtained directly from the mobile phone operator through an agreement between the Data Science Institute and Telef\'onica R\&D. This mobile phone operator retains ownership of these data and imposes standard provisions to their sharing and access which guarantee privacy. Anonymized datasets are available from Telef\'onica R\&D Chile (\url{http://www.tidchile.cl}) for researchers who meet the criteria for access to confidential data. Other datasets used in this study are either derived from mobile records, or publicly available.

\bibliographystyle{bmc-mathphys}
\bibliography{references}

\end{document}